\documentclass[aps,prl,reprint,amsmath,amssymb,superscriptaddress]{revtex4-2}

\usepackage{graphicx}
\usepackage{dcolumn}
\usepackage{bm}
\usepackage{verbatim}
\usepackage{multirow}
\usepackage{xcolor}
\usepackage{tabularx}

\begin{document}

\preprint{APS/123-QED}

\title{Wide Effective Work-Function Tuning of Al/SiO$_2$/Si Junction Achieved with Graphene Interlayer at Al/SiO$_2$ Interface}

\author{Wonho Song}
\affiliation{Department of Physics, Ulsan National Institute of Science and Technology (UNIST), Ulsan 44919, Republic of Korea}

\author{Jung-Yong Lee}
\affiliation{Korea Development Bank, Seoul 07242, Republic of Korea}

\author{Junhyung Kim}
\affiliation{Department of Physics, Ulsan National Institute of Science and Technology (UNIST), Ulsan 44919, Republic of Korea}

\author{Jinyoung Park}
\affiliation{Department of Physics, Ulsan National Institute of Science and Technology (UNIST), Ulsan 44919, Republic of Korea}

\author{Jaehyeong Jo}
\affiliation{Department of Physics, Ulsan National Institute of Science and Technology (UNIST), Ulsan 44919, Republic of Korea}

\author{Eunseok Hyun}
\affiliation{Department of Physics, Ulsan National Institute of Science and Technology (UNIST), Ulsan 44919, Republic of Korea}

\author{Jiwan Kim}
\affiliation{Department of Physics, Ulsan National Institute of Science and Technology (UNIST), Ulsan 44919, Republic of Korea}

\author{Daejin Eom}
\affiliation{Korea Research Institute of Standards and Science (KRISS), Daejeon 34113, Republic of Korea}

\author{Gahyun Choi}
\email{ghchoi@kriss.re.kr}
\affiliation{Korea Research Institute of Standards and Science (KRISS), Daejeon 34113, Republic of Korea}

\author{Kibog Park}
\email{kibogpark@unist.ac.kr}
\affiliation{Department of Physics, Ulsan National Institute of Science and Technology (UNIST), Ulsan 44919, Republic of Korea}
\affiliation{Department of Electrical Engineering, Ulsan National Institute of Science and Technology (UNIST), Ulsan 44919, Republic of Korea}

\begin{abstract}

The effective work-function of metal electrode is one of the major factors to determine the threshold voltage of metal/oxide/semiconductor junction. In this work, we demonstrate experimentally that the effective work-function of Aluminum (Al) electrode in Al/SiO$_2$/n-Si junction increases significantly by $\sim$1.04 eV with the graphene interlayer inserted at Al/SiO$_2$ interface. We also provide the device-physical analysis of solving Poisson equation when the flat-band voltage is applied to the junction, supporting that the wide tuning of Al effective work-function originates from the electrical dipole layer formed by the overlap of electron orbitals between Al and graphene layer. Our work suggests the feasibility of constructing the dual-metal gate CMOS circuitry just by using Al electrodes with area-specific underlying graphene interlayer.

\end{abstract}

\maketitle

The metal/oxide/semiconductor (MOS) structure has been used as an essential element in electronic device applications. It itself can function as a voltage-dependent variable capacitor. However, the most prominent usage is to work as the gate stack and current channel of MOS field effect transistor (MOSFET). In the real-world applications, both n- and p-channel MOSFET are used in combinatorial manners for minimizing the power consumption during device operation, bearing the name of complementary MOS (CMOS). There has been a tremendous amount of researches for improving and optimizing the operational characteristics of MOSFET \cite{MOSFET}. In particular, the threshold voltages to form inversion channels in n- and p-channel MOSFET are strongly preferred to be symmetric from zero, meaning the same magnitude but opposite polarity, as much as possible \cite{MOS2,Dual}. These threshold voltages depend primarily on the work-function of gate metal electrode. More practical quantity for the MOS structure is the so-called effective work-function which is defined as the energy difference between Fermi-level of metal and vacuum level of semiconductor in the flat-band situation \cite{EWF,EWF2}. Thus, the various layer-structural and chemical methods modulating the effective work-function of gate metal have been adopted including metal interdiffusion \cite{MOS6}, post treatment of insulating layer \cite{MOS,MOS2,MOS3,Dual}, chemical reaction of metal and insulator \cite{MOS4}, and multiple oxide layer \cite {MOS5}. Recently, it has been reported that the work-function of a metal film can be modulated when the metal film being in contact with a graphene monolayer \cite{DFT,NP}. In this letter, we demonstrate experimentally that the flat-band voltage of metal/SiO$_2$/n-Si junction, interlocked with the effective work-function of gate metal, can change significantly with a graphene interlayer inserted at metal/SiO$_2$ interface. We also show that the effective work-function change of metal/graphene stack can stem from the electric dipole layer formed between metal and graphene by solving Poisson equation to obtain the electron energy band profile of junction \cite{DFT,NP}. Most relevantly, the effective work-function of Al/graphene stack is found to increase quite a lot compared with Al, low work-function ($\sim$4.08 eV, \cite{Al}) metal suitable for n-channel MOSFET gate electrode, enough to be used as a gate electrode of p-channel MOSFET. This implies that it will be possible to form the gate electrodes for both n- and p-channel MOSFET just with the Al metalization process \cite{MOS2,Dual,Dual2}.

\begin{figure}
\includegraphics{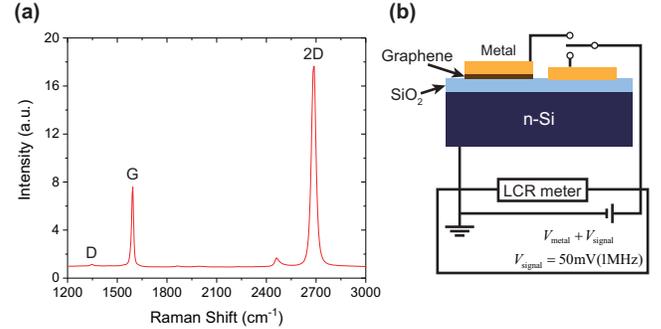}
\caption{\label{F1}(a) Raman spectrum measured on transferred monolayer graphene with D, G, and 2D peaks being indicated. (b) Schematic view of MOS and MGOS junctions where the capacitance-voltage measurement configuration using LCR meter is drawn. The AC signal voltage $V_{\rm signal}$ is applied with the \textit{rms} magnitude of 50 mV and the frequency of 1 MHz while DC bias $V_{\rm metal}$ is applied.}
\end{figure}

\begin{figure}
\includegraphics{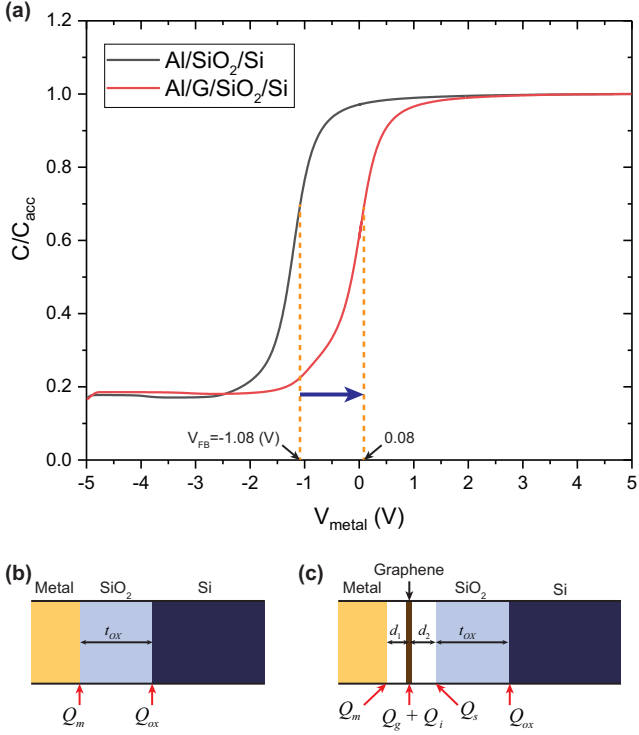}
\caption{\label{F2} (a) Capacitance ($C$), normalized to accumulation capacitance ($C_{acc}$), measured as a function of voltage applied on metal ($V_{\rm metal}$) for Al/SiO$_2$/Si and Al/graphene/SiO$_2$/Si junctions. The measurements were performed on multiple junctions for both cases and one specific measurement outcome is shown for each case. The $V_{FB}$ of each junction is indicated in the horizontal axis. Al/graphene electrode is written shortly as Al/G. Charge distributions of (b) Al/SiO$_2$/n-Si and (c) Al/graphene/SiO$_2$/n-Si junctions at $V_{\rm metal}=V_{FB}$.}
\end{figure}

The metal/graphene/oxide/semiconductor (MGOS) junctions for investigating the effective work-function modulation were prepared as follows. A $\sim$30 nm thick SiO$_2$ layer was grown on an n-type Si wafer (donor concentration $N_{D} \sim 3\times10^{15}$ cm$^{-3}$) by using the dry oxidation and the SiO$_2$/Si substrate was cleaned in acetone and methanol each for 5 minutes sequentially. Then, a graphene layer grown with chemical vapor deposition (CVD), purchased from the Graphene Supermarket Inc., was transferred partly onto the substrate by using the semi-dry transfer method reported previously \cite{Dry}. After transferring the graphene layer, Raman spectroscopy was performed to evaluate the quality of graphene layer. As shown in Fig. \ref{F1}(a), the ratios of D peak to G peak and 2D peak to G peak are estimated to be $\sim$0.03 and $\sim$2.52 respectively. These indicate that the transferred graphene is a monolayer having a negligible number of defects \cite{Raman,Raman2}. To form the MOS and MGOS structures under the same metallization condition, circular metal electrodes with their diameter of $\sim$500 $\mu$m were deposited on the SiO$_2$/Si and graphene/SiO$_2$/Si areas simultaneously through a shadow mask. Then, the sample was exposed to O$_2$ plasma for isolating electrically individual metal electrodes underlined with the graphene interlayer. We first used aluminum (Al) electrodes to form the MOS and MGOS structures which were deposited by using thermal evaporation. The interaction of Al with graphene has been reported to be relatively weak enough to preserve the electronic structure of graphene around the Dirac point, meaning that the so-called physisorption occurs at the interface \cite{DFT,NP}. Hence, it is possible for us to use the linear density-of-states relation of graphene in calculating the energy band profile across the MGOS junction \cite{Qg,GaSe}.

After the sample preparation, we performed the capacitance-voltage (C-V) measurements to obtain the flat-band voltages ($V_{FB}$) of Al/SiO$_2$/Si and Al/graphene/SiO$_2$/Si junctions by using the Agilent E4980A LCR meter. The small AC voltage for probing the capacitive response of junction was set to be 50 mV (rms) at the frequency of 1 MHz and the DC bias voltage ($V_{\rm metal}$) applied onto the metal electrode varies from -5 to 5 V as illustrated in Fig. \ref{F1}(b). Figure \ref{F2}(a) shows the measured C-V curves for Al electrode where the typical capacitance characteristics of p-channel MOS are observed. Here, the C-V curve for Al electrode is found to be shifted to the right with the graphene interlayer, implying that the effective work-function of Al electrode increases accordingly.

By estimating the flat-band capacitance in the measured C-V curve, we extracted the $V_{FB}$ of each junction \cite{VFB3}. The average  $V_{FB}$ of Al electrode measured over three different junctions for each of Al/SiO$_2$/Si and Al/graphene/SiO$_2$/Si junctions was found to be -0.98 V and 0.06 V, respectively, indicating that the $V_{FB}$ shift is quite significant to be 1.04 V. From the extracted $V_{FB}$, the effective work-function of metal electrode can be obtained with the simple MOS capacitance formalism assuming the existence of fixed charges ($Q_{ox}$) in the SiO$_2$ layer. If $V_{\rm metal}$ is equal to the $V_{FB}$, the depletion or inversion charges on the semiconductor side should be zero. Then, the $Q_{ox}$ including bulk electron traps and ionic impurities can be obtained from eq. 1 below reflecting the charge distribution shown in Fig. \ref{F2}(b) \cite{Eq,Wm}.
\begin{flalign}
\phi_m=\chi_{\rm s}+V_n+V_{FB}+\frac{Q_{ox}t_{ox}}{\epsilon_{ox}\epsilon_0}
\end{flalign}
where $\phi_m$ is the metal work-function, $\chi_s$ is the electron affinity of Si, $V_n$ is the energy difference between the conduction band minimum and the Fermi level of bulk n-Si, $t_{ox}$ is the thickness of SiO$_2$ layer, $\epsilon_0$ is the vacuum permittivity and $\epsilon_{ox}$ is the dielectric constant of SiO$_2$. $V_n$ is calculated with $(k_BT/q){\rm ln}(N_C/N_D)$ where $k_B$ is the Boltzmann constant, $T$ is the temperature, $q$ is the elementary charge, and $N_C$ is the effective density-of-states for the conduction band of Si. $Q_{ox}$ is assumed to be located effectively at the SiO$_2$/Si interface \cite{VFB2,VFB3,MOS2,MOS3}. It is also reasonable to assume that $Q_{ox}$ is the same for MOS and MGOS junctions because both junctions are formed on the identical SiO$_2$/Si substrate.

As described before, the effective metal work-function $\phi_m^e$ of MOS capacitor represents the energy difference between Fermi-level of metal and vacuum level of semiconductor in the flat-band situation \cite{EWF,EWF2}. Then, the flat-band voltage should be equal to the difference of $\phi_m^e$ and work-function of semiconductor $\phi_s$. That is, $V_{FB}=\phi_m^e-\phi_{s}$. Here, $\phi_{s}$ can be expressed as a sum of $\chi_s$ and $V_n$. Hence, the effective metal work-function becomes $\phi_m^e=\chi_s+V_n+V_{FB}$. The effective work-function of Al electrode is obtained to be $\sim$3.31 eV and $\sim$4.35 eV for Al/SiO$_2$/Si and Al/graphene/SiO$_2$/Si junctions, respectively. The increase of effective work-function with the graphene interlayer, amounting to over 1.00 eV, is quite significant as mentioned previously. Since the intrinsic work-function of graphene ($\phi_g$) is $\sim$4.50 eV \cite{Wg} larger just by $\sim$0.42 eV compared with Al, it seems not feasible to allocate the graphene work-function as the origin for such a large increase of effective work-function. We need to explore other physical mechanisms to explain this wide tuning in the effective work-function of Al electrode. Similar phenomena have been reported in several previous researches including metal/graphene/GaAs junctions \cite{NP} and metal/graphene/Ge junctions \cite{Ge}. One plausible way is to adopt the existence of an electric dipole layer formed at the metal/graphene interface stemming from the off-centric distribution of the overlapped electron wave functions between metal and graphene layers \cite{DFT,NP}. According to the density functional theory (DFT) calculation done by Khomyakov \textit{et.al} \cite{DFT}, the electrons mediating the bonding between metal and graphene are distributed more closely to the metal side for all physisorbed metals. This implies that an electric dipole layer forms at the metal/graphene interface with its direction pointing from metal to graphene. If denoting $Q_i$ as the charge on the graphene side of the interface dipole layer, $Q_i$ should be positive for all physisorbed metals. However, the polarity of $Q_i$ was found to be inverted when metals with relatively low work-function were contacted to graphene, observed in metal/graphene/GaAs junctions containing local patches of weak Fermi-level pinning. \cite{NP}. By relying on the observation in metal/graphene/GaAs junctions, the $Q_i$ is expected to be negative for the Al/graphene contact as well.

For more quantitative investigation on how the electric dipole layer at Al/graphene interface influences the measured C-V curves, we have performed the analytical electrostatic modeling to obtain the electron energy band profile across both MOS and MGOS junctions under the flat-band conditions. Figure \ref{F2}(b) and(c) show the charge distributions used in the modeling for MOS and MGOS junctions, respectively. Here, $Q_m$ is the metal surface charge and $Q_g$ is the free charge in the graphene layer reflecting the charge carrier transfer between metal and graphene layers due to the Fermi-level difference between them. The gap between metal and graphene layer $d_1$ is chosen to be $\sim$3.3 {\AA} reported previously with the DFT calculation \cite{DFT}. The gap between graphene and SiO$_2$ layer $d_2$ is assumed to be $\sim$5.0 {\AA}, somewhat larger than the theoretically-predicted value $\sim$3.0 {\AA} \cite{DFT2}, by considering the unavoidable existence of various wrinkles formed during the graphene transfer process. The wrinkles will reduce the overall flatness of transferred graphene, making the graphene and SiO$_2$ layers further apart from each other and increasing the average spacing between the two layers for the entire junction \cite{Gap}. It is well-known that the graphene layer transferred on the SiO$_2$ surface is p-type doped due to the electron transfer from the graphene layer to the surface states of SiO$_2$ \cite{p-doping}. The negative charge density residing on the SiO$_2$ surface is labeled as $Q_s$ which originates from the O-dangling bond states of SiO$_2$. The energy of O-dangling bond state is reported to be $\sim$1.5 eV below the Dirac point of graphene \cite{Ostate}. Hence, it is reasonably assumed that the O-dangling bond states are completely filled with electrons and $Q_s$ stays constant regardless of what metal being used as the gate electrode and the range of gate voltage used in this work.

\begin{figure}
\includegraphics{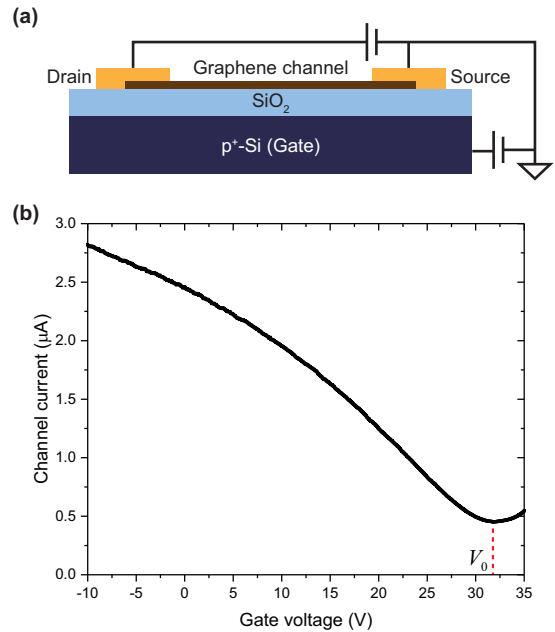}
\caption{\label{F3} (a) Schematic cross-sectional view of GFET and (b) measured source-drain current vs. gate voltage.}
\end{figure}

Based on the charge distributions shown in Fig. \ref{F2}, we first calculate the electrostatic potential $U(x)$ by solving the one-dimensional Poisson equation $d^2U/dx^2 = - \rho / \epsilon_r\epsilon_0$ where $\epsilon_r$ is the dielectric constant of either SiO$_2$ ($\epsilon_{ox}$) or Si ($\epsilon_s$) \cite{Eq,GaSe}. Then, we derived the analytical expression of $Q_i$ as shown in eq. (2) below (for more detailed description for the electrostatic potential calculation including boundary conditions and charge densities, see Supplemental Material \cite{supp}). The material parameters used in the calculation such as dielectric constant, metal work-function, and band gap were referred to the literatures and are listed in Table \ref{T1} \cite{DFT,Wg,Al,Pt,SiO2,Si}.
\renewcommand{\arraystretch}{1.3}
\begin{table*}[]
\begin{ruledtabular}
\begin{tabular}{ccccccccc}
$\phi_m$ (Al) & $\phi_m$ (Pt) & $\chi_s$ & $\epsilon_{ox}$ & $\epsilon_s$ & $N_C$ & $\phi_g$ & $v_F$ \\ \hline
4.08 eV \cite{Al} & 5.65 eV \cite{Pt} & 4.05 eV \cite{Si} & 3.9 \cite{SiO2} & 11.7 \cite{Si} & $2.82\times 10^{19}$ cm$^{-3}$ \cite{Si} & 4.50 eV \cite{Wg} & $10^6$ m/s \cite{Wg} \\
\end{tabular}
\caption{\label{T1} {Material parameters used in analytical electrostatic modeling}}
\end{ruledtabular}
\end{table*}
\renewcommand{\arraystretch}{1}
\begin{flalign}
Q_i=-Q_s-Q_{ox}-\frac{\epsilon_0}{d_1}\left(V_{FB}-\phi_m+\chi_s+V_n \right) \nonumber
\\-\frac{d_2}{d_1}\left( Q_s+Q_{ox} \right)-\frac{t_{ox}}{d_1\epsilon_{\rm SiO_2}}Q_{ox}-Q_g
\end{flalign}
Here, the graphene free charge is quantified as $Q_g=(q\Delta E_F |\Delta E_F|)/(\pi \hbar^2 v_F^2)$ where $\Delta E_F$ is the difference between Fermi-level and Dirac point of graphene layer and $v_F$ is the Fermi-velocity of graphene \cite{Qg,GaSe}. $Q_g$ can be determined by substituting $\Delta E_F=-\phi_g+ V_{FB}+\chi_s+V_n + \frac {d_2} {\epsilon_0} (Q_s+Q_{ox}) + \frac {t_{ox}} {\epsilon_{ox} \epsilon_0} Q_{ox}$ derived from the calculated $U(x)$. Here, $\phi_g$ is the intrinsic work-function of graphene chosen to be $4.50$ eV. Then, the only undetermined quantity is $Q_s$ and it can be obtained experimentally from the transfer curve (source-drain current vs. gate voltage) of graphene field effect transistor (GFET) \cite{Dry}. A GFET was fabricated on a highly p-doped Si substrate with a 100 nm thermally-grown SiO$_2$ layer on top. First, a graphene layer was transferred onto the substrate, then the source and drain electrodes were formed by e-beam evaporating Ti/Au film stacks through a shadow mask. After that, the graphene channel was defined with photolithography and subsequent O$_2$ plasma etching processes. Finally, the remaining photoresist was removed with acetone. The schematic illustration of fabricated GFET is shown in Fig. \ref{F3}(a). The transfer curve of fabricated GFET shown in Fig. \ref{F3}(b) was obtained by sweeping the gate voltage from -10 to 35 V with the drain voltage of 0.1 V. In Figure \ref{F3}(b), the gate voltage inducing the minimum channel current (Charge Neutrality Point, CNP) is labeled as $V_0$ and its average value is obtained to be $\sim$32.3 V from the measurements on three GFETs. At $V_g$ = 0 V, the net charge density in the graphene channel, amounting to its initial p-type (hole) doping charge density, is supposed to be equal to the surface charge density on the SiO$_2$ layer ($Q_s$). When the gate bias reaches the CNP ($V_g=V_0$), the net charge density in the graphene channel becomes nearly zero. Then, the $Q_s$ will be compensated by the dielectric charge density induced on the SiO$_2$ surface, leading to $Q_s=\epsilon_{ox}\epsilon_0V_0/(100 ~{\rm nm})$. From the known or experimentally-obtained values of the related parameters, it is obtained that $Q_s=-1.12\times10^{-6}$ C/cm$^2$. Subsequently, the $Q_i$ of Al/graphene interface is calculated to be $-2.85\times10^{-7}$ C/cm$^2$ from eq. (2). Here, it is noted that the $Q_i$ is found to be negative consistently with the previous experiments on Al/graphene/GaAs junctions \cite{NP}.

Based on the calculation above, the flat-band diagrams for Al/SiO$_2$/Si and Al/graphene/SiO$_2$/Si junctions are obtained as shown in Fig. \ref{F4} and the values of relevant parameters are summarized in the Supplemental Material \cite{supp}. One important point to make is that the increase of effective work-function ($\sim$1.04 eV) is very similar to the band gap of Si ($\sim$1.12 eV) and the work-function of Al electrode ($\sim$4.08 eV) is very similar to the electron affinity of Si ($\sim$4.05 eV). This implies that the Al electrode suitable for n-channel MOSFET due to its small work-function can also be used for p-channel MOSFET with a graphene interlayer. Hence, it will be possible to realize the dual-metal gate CMOS system with Al electrodes \cite{MOS2,Dual,Dual2}.

\begin{figure}
\includegraphics{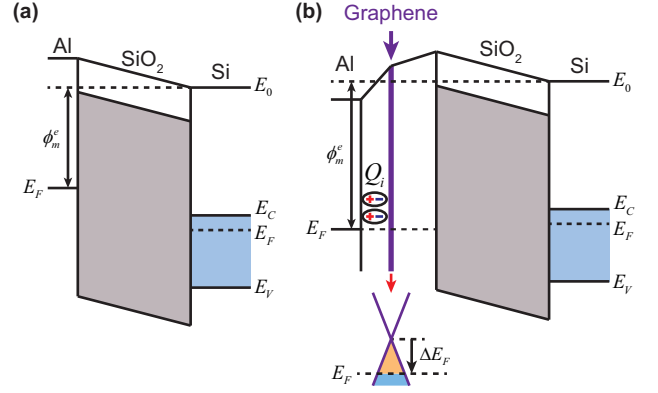}
\caption{\label{F4} Energy band profiles for (a) Al/SiO$_2$/Si and (b) Al/graphene/SiO$_2$/Si junctions under the flat-band voltage condition. Here, $E_0$ is the vacuum level, $E_C$ is the conduction band minimum, $E_V$ is the valance band maximum, and $E_F$ is the Fermi-level.}
\end{figure}

\begin{figure}
\includegraphics{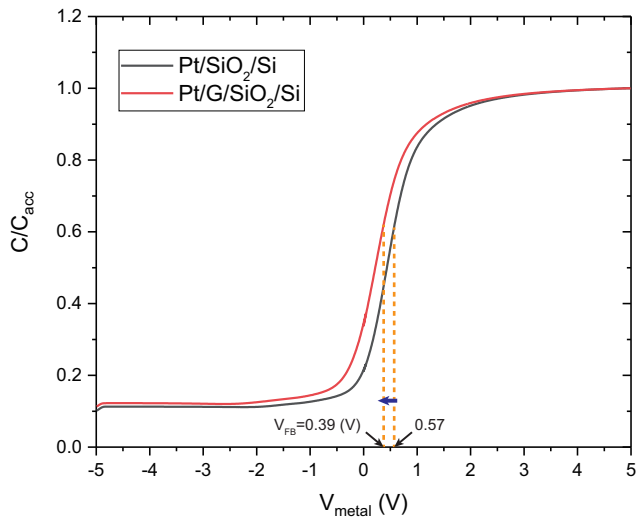}
\caption{\label{F5} Capacitance ($C$), normalized to accumulation capacitance ($C_{acc}$), measured as a function of voltage applied on metal ($V_{\rm metal}$) for Pt/SiO$_2$/Si and Pt/graphene/SiO$_2$/Si junctions. The measurements were performed on multiple junctions for both cases and one specific measurement outcome is shown for each case. The $V_{FB}$ of each junction is indicated in the horizontal axis.}
\end{figure}

For comparing with a metal electrode of high work-function, MOS and MGOS junctions using Pt electrodes were also prepared by following the fabrication procedures identical to the Al electrode case. The work-function of Pt is $\sim$5.65 eV, higher than that of graphene ($\sim$4.50 eV). Similarly to Al, the bonding between Pt and graphene is reported to be weak (physisorption). From the C-V measurements shown in the Fig. \ref{F5}, the effective work-function of Pt electrode is found to decrease with the graphene interlayer. Specifically, the averaged effective work-function is measured to decrease from $\sim$4.85 to $\sim$4.66 eV manifested in the relatively small $V_{FB}$ shift from $\sim$0.57 to $\sim$0.37 V \cite{Pt}. The $Q_i$ for Pt/graphene interface was calculated to produce the shift of effective work-function extracted from C-V measurements and it is listed in Table \ref{T2}. The effective work-function decrease ($\sim$0.19 eV) with the aid of positive $Q_i$ seems consistent with the previous work for Pt/graphene/GaAs junction where the Schottky barrier of junction was found to decrease substantially with the graphene interlayer. \cite{NP}. The synergetic implication of the measurements with the two metals (Al, Pt) is that the effective work-function shifts observed for them cannot be explained altogether just by considering the free carrier doping of graphene layer. More concretely, if we don't include the $Q_i$, the other varying parameter $d_2$ needs to be different by an unreasonably large amount between the two metals in order to generate the shift of effective work-function correctly for both metals. Hence, it seems inevitable to adopt the existence of $Q_i$ for analyzing the measured data properly. The further detailed discussion on this point can be found in the Supplemental Material \cite{supp}.

\renewcommand{\arraystretch}{1.3}
\begin{table}[]
\begin{ruledtabular}
\begin{tabular}{lcccc}
 & Al & Al/G & Pt & Pt/G \\ \hline
$V_{FB}$ (V) & -0.98 & 0.06 & 0.57 & 0.37 \\
$\phi_m^e$ (eV) & 3.31 & 4.35 & 4.85 & 4.66 \\
$Q_i$ (10$^{-7}$ C/cm$^2$) & - & -2.85 & - & 15.98 \\
\end{tabular}
\caption{\label{T2} Averaged values of flat-band voltage, effective work-function, and interaction dipole charge obtained for Al and Pt electrodes \textit{with} and \textit{without} the graphene interlayer inserted at the metal/SiO$_2$ interface of metal/SiO$_2$/Si junction where $Q_i$ value can be obtained only for metal/graphene/SiO$_2$/Si junctions.}
\end{ruledtabular}
\end{table}
\renewcommand{\arraystretch}{1}

In conclusion, we have observed a significant increase of effective work-function ($\sim$1.04 eV) in Al/graphene/SiO$_2$/n-Si junction in comparison with Al/SiO$_2$/n-Si junction. The wide tuning of effective work-function was observed in C-V measurements and the analytical calculation of solving one-dimensional Poisson equation at flat-band voltage was performed to figure out its physical origin. In the calculation, an electric dipole layer was adopted to form between metal and graphene, originating from the overlap of electron orbitals. A similar effective work-function shift was observed also with Pt electrode. This time, the effective work-function was found to decrease by $\sim$0.19 eV. In order to account for the observed shifts of effective work-function within the reasonable range of the spacing between graphene and SiO$_2$ layer, the interaction dipole layer is found to have its negative side toward the graphene layer for Al while the polarity is flipped over (the positive side toward the graphene layer) for Pt. If considering the commonness and compatibility with the existing CMOS processes of Al as additional advantages, our work suggests that it will be possible to construct the dual-metal gate CMOS circuitry by using the Al electrodes with area-specific underlying graphene interlayers in cost-effective and reliable manners \cite{MOS2,Dual,Dual2}.

\begin{acknowledgments}

This work was supported by the National Research Foundation of Korea (NRF) funded by Ministry of Science and ICT (2020M3F3A2A02082437, 2019R1A5A1027055, 2020K2A9A1A06103641, {2018H1A 2A1062406}). This work was also supported by Samsung Research Funding \& Incubation Center of Samsung Electronics under Project Number SRFC-TA1903-02 and has benefited from the use of the facilities at UNIST Central Research Facilities.

\end{acknowledgments}


\begin{thebibliography}{99}

\bibitem{MOSFET}
R. K. Ratnesh, A. Goel, G. Kaushik, H. Garg, Chandan, M. Singh, and B. Prasad, Mater. Sci. Semicond. Process. {\bf 134}, 106002 (2021).

\bibitem{MOS2}
C. Zhao and J. Xiang, Appl. Sci. {\bf 9}, 2388 (2019).

\bibitem{Dual}
A. Fet, V. Häublein, A. J. Bauer, H. Ryssel, and L. Frey, Appl. Phys. Lett. {\bf 96}, 053506 (2010).

\bibitem{EWF}
H. Zhu and R. Ramprasad, Phys. Rev. B {\bf 83}, 081416(R) (2011).

\bibitem{EWF2}
M. Xueli, Y. Hong, W. Wenwu, Y. Huaxiang, Z. Huilong, Z. Chao, C. Dapeng, and Y. Tianchun, J. Semicond. {\bf 35}, 096001 (2014).

\bibitem{MOS6}
I. Polishchuk, P. Ranade, T.-J. King, and C. Hu, IEEE Electron Device Lett. {\bf 22}, 444 (2001).

\bibitem{MOS}
J. Tao, C. Z. Zhao, C. Zhao, P. Taechakumput, M. Werner, S. Taylor, and P. R. Chalker, Materials {\bf 5}, 1005 (2012).

\bibitem{MOS3}
M. Ťapajna, K. Hušeková, J. P. Espinos, L. Harmatha, and K. Fröhlich, Mater. Sci. Semicond. Process. {\bf 9}, 969 (2006).

\bibitem{MOS4}
J. Lin, Y. Y. Gomeniuk, S. Monaghan, I. M. Povey, K. Cherkaoui, É. O'Connor, P. Máire, and P. K. Hurley, J. Appl. Phys. {\bf 114}, 144105 (2013).

\bibitem{MOS5}
X. J. Liu, L. Zhu, M. Y. Gao, X. F. Li, Z. Y. Cao, H. F. Zhai, A. Li, and D. Wu, Appl. Surf. Sci. {\bf 289}, 332 (2014).

\bibitem{DFT}
P. A. Khomyakov, G. Giovannetti, P. C. Rusu, G. Brocks, J. van den Brink, and P. J. Kelly, Phys. Rev. B {\bf 79}, 195425 (2009).

\bibitem{NP}
H. H. Yoon, W. Song, S. Jung, J. Kim, K. Mo, G. Choi, H. Y. Jeong, J. H. Lee, and K. Park, ACS Appl. Mater. Interfaces {\bf 11}, 47182 (2019).

\bibitem{Al}
P. A. Tipler and R. A. Llewellyn, {\it Modern Physics} (W. H. Freeman and Company, New York, 2003), Vol. 5, pp. 130.

\bibitem{Dual2}
Y.-C. Yeo, Q. Lu, P. Ranade, H. Takeuchi, K. J. Yang, I. Polishchuk, T.-J. King, C. Hu, S. C. Song, H. F. Luan, and D.-L. Kwong, IEEE Electron Device Lett. {\bf 22}, 227 (2001)

\bibitem{Dry}
S. Jung, H. H. Yoon, H. Jin, K. Mo, G. Choi, J. Lee, H. Park and K. Park, J. Appl. Phys. {\bf 125}, 184302 (2019).

\bibitem{Raman}
A. C. Ferrari, Solid State Commun. {\bf 143}, 47 (2007).

\bibitem{Raman2}
A. Das, B. Chakraborty, and A. K. Sood, Bull. Mater. Sci. {\bf 31}, 579 (2008)

\bibitem{Qg}
A. H. Castro Neto, F. Guinea, N. M. R. Peres, K. S. Novoselov, and A. K. Geim, Rev. Mod. Phys. {\bf 81}, 109 (2009).

\bibitem{GaSe}
W. Kim, C. Li, F. A. Chaves, D. Jiménez, R. D. Rodriguez, J. Susoma, M. A. Fenner, H. Lipsanen, and J. Riikonen, Adv. Mater. {\bf 28}, 1845 (2016).

\bibitem{VFB3}
K. Piskorski and H. M. Przewlocki, “The methods to determine flat-band voltage VFB in semiconductor of a MOS structure,” in MIPRO, Opatija, Croatia, 24–28 May (2010).

\bibitem{Eq}
S. M. Sze and K. K. Ng, {\it Physics of semiconductor devices} (John wiley \& sons, New York, 2006).

\bibitem{Wm}
D. K. Schroder, {\it Semiconductor Material and Device Characterization} (John Wiley \& Sons, Inc., Hoboken, NJ, USA, 2015), Vol. 3, p. 336.

\bibitem{VFB2}
C. Cobianu, F. Nastase, N. Dumbravescu, O. Buiu, B. Serban, M. Danila, R. Gavrila, O. Ionescu, and C. Romanitan, Rom. J. Inf. Sci. Technol. {\bf 22}, 41 (2019).

\bibitem{Wg}
Y. J. Yu, Y. Zhao, S. Ryu, L. E. Brus, K. S. Kim, and P. Kim, Nano Lett. {\bf 9}, 3430 (2009).

\bibitem{Ge}
S. H. C. Baek, Y. J. Seo, J. G. Oh, M. G. A. Park, J. H. Bong, S. J. Yoon, and S. H. Lee, Appl. Phys. Lett. {\bf 105}, 073508 (2014).

\bibitem{DFT2}
W. Gao, P. Xiao, G. Henkelman, K. M. Liechti, and R. Huang, J. Phys. D {\bf 47}, 255301 (2014).

\bibitem{Gap}
H. J. Zhong, Z. H. Liu, L. Shi, G. Z. Xu, Y. M. Fan, Z. L. Huang, J. F. Wang, G. Q. Ren, and K. Xu, Appl. Phys. Lett. {\bf 104}, 212101 (2014).

\bibitem{p-doping}
X. F. Fan, W. T. Zheng, V. Chihaia, Z. X. Shen, and J.-L. Kuo, J. Phys.: Condens. Matter {\bf 24}, 305004 (2012).

\bibitem{Ostate}
H. J. Sung, D. H. Choe, and K. J. Chang, New J. Phys. {\bf 16}, 113055 (2014).

\bibitem{supp}
See the Supplemental Material for details on formalism for analytical electrostatic modeling and the effect of interaction dipole layer.

\bibitem{Pt}
H. B. Michaelson, J. Appl. Phys. {\bf 48}, 4729 (1977).

\bibitem{SiO2}
C. M. Tanner, Y. C. Perng, C. Frewin, S. E. Saddow, and J. P. Chang, Appl. Phys. Lett. {\bf 91}, 203510 (2007).

\bibitem{Si}
B. J. Baliga, {\it Fundamentals of Power Semiconductor Devices} (Springer, New York, 2010), pp. 24.

\end{thebibliography}
\end{document}


\title{Supplemental Material:\\Wide Effective Work-Function Tuning of Al/SiO$_2$/Si Junction Achieved with Graphene Interlayer at Al/SiO$_2$ Interface}

\author{Wonho Song}
\affiliation{Department of Physics, Ulsan National Institute of Science and Technology (UNIST), Ulsan 44919, Republic of Korea}

\author{Jung-Yong Lee}
\affiliation{Korea Development Bank, Seoul 07242, Republic of Korea}

\author{Junhyung Kim}
\affiliation{Department of Physics, Ulsan National Institute of Science and Technology (UNIST), Ulsan 44919, Republic of Korea}

\author{Jinyoung Park}
\affiliation{Department of Physics, Ulsan National Institute of Science and Technology (UNIST), Ulsan 44919, Republic of Korea}

\author{Jaehyeong Jo}
\affiliation{Department of Physics, Ulsan National Institute of Science and Technology (UNIST), Ulsan 44919, Republic of Korea}

\author{Eunseok Hyun}
\affiliation{Department of Physics, Ulsan National Institute of Science and Technology (UNIST), Ulsan 44919, Republic of Korea}

\author{Jiwan Kim}
\affiliation{Department of Physics, Ulsan National Institute of Science and Technology (UNIST), Ulsan 44919, Republic of Korea}

\author{Daejin Eom}
\affiliation{Korea Research Institute of Standards and Science (KRISS), Daejeon 34113,Republic of Korea}

\author{Gahyun Choi}
\email{ghchoi@kriss.re.kr}
\affiliation{Korea Research Institute of Standards and Science (KRISS), Daejeon 34113,Republic of Korea}

\author{Kibog Park}
\email{kibogpark@unist.ac.kr}
\affiliation{Department of Physics, Ulsan National Institute of Science and Technology (UNIST), Ulsan 44919, Republic of Korea}
\affiliation{Department of Electrical Engineering, Ulsan National Institute of Science and Technology (UNIST), Ulsan 44919, Republic of Korea}

\maketitle

\section{Formalism for Analytical Electrostatic Modeling}

\begin{figure}
\sfigone
\includegraphics{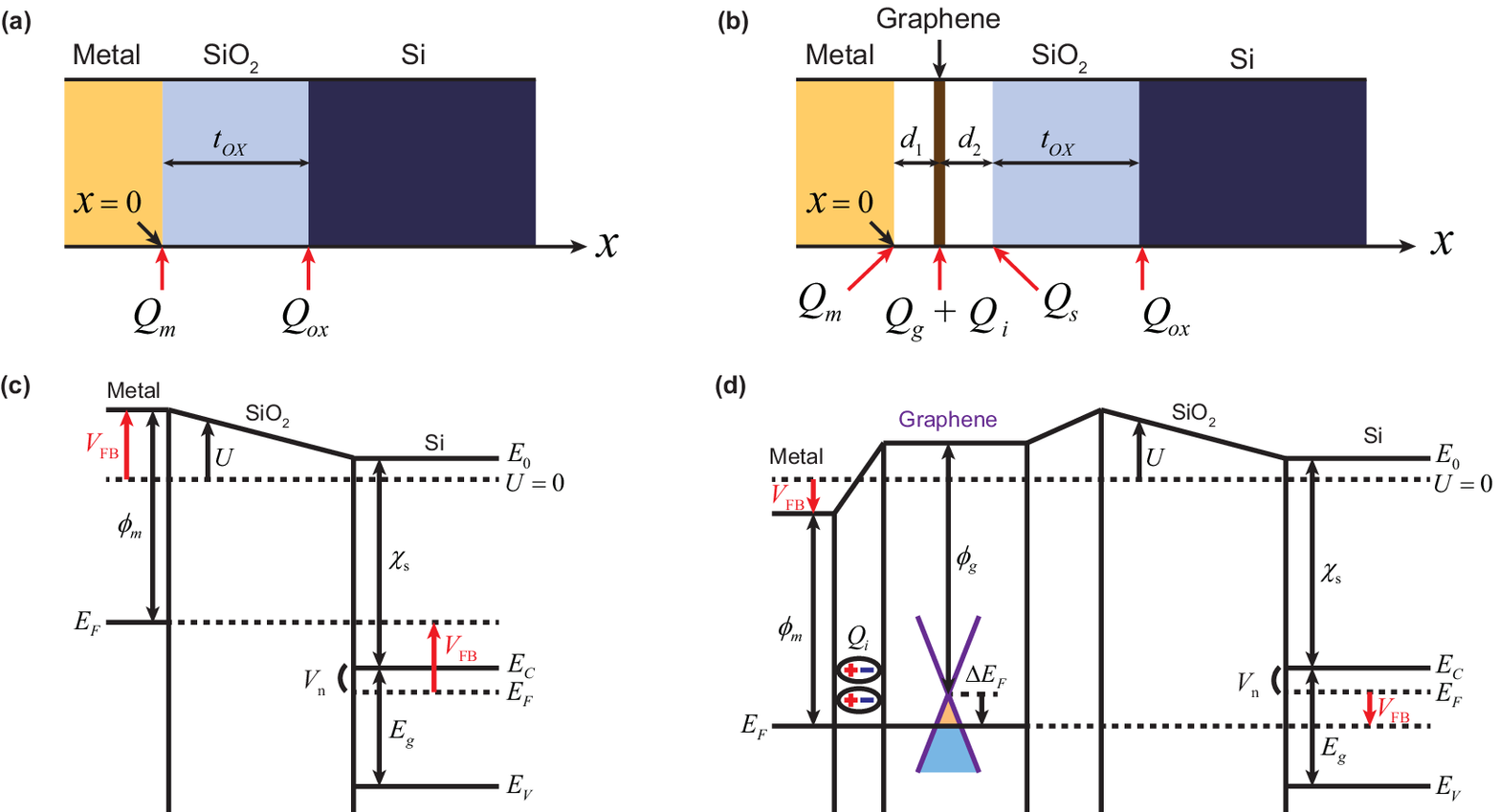}
\caption{\label{SF1} Spatial charge distributions and flat-band diagrams for metal/SiO$_2$/n-Si ((a), (c)) and metal/graphene/SiO$_2$/n-Si ((b), (d)) junctions. The band diagrams were drawn for the case of Al metal electrodes.}
\end{figure}

The flat-band diagram of metal/oxide/semiconductor (MOS) and  metal/graphene/oxide/semiconductor (MGOS) junctions are calculated by solving one-dimensional Poisson equation with the charge distributions shown in Fig. \ref{SF1}(a) and (b). The parameters used in the calculation are defined as follows, $Q_m$: Metal surface charge density, $Q_{ox}$: Fixed oxide charge density at the SiO$_2$/n-Si interface, $Q_g$: Doping charge density of graphene, $Q_i$: Dipole charge density formed by interaction between metal and graphene, $Q_s$: Trap charge density on the graphene-side SiO$_2$ surface making the graphene layer p-type doped, $t_{ox}$($=30$ nm): Thickness of SiO$_2$ layer, $d_1$($=3.3$ \AA ~\cite{DFT}): Gap between metal and graphene, $d_2$($=5$ \AA): Gap between graphene and SiO$_2$, $\phi_m$(= 4.08 eV for Al \cite{Al}, 5.65 eV for Pt \cite{Pt}): Metal work-function, $\chi_s$($=4.05$ eV \cite{Si}): Electron affinity of Si, $V_n$($=(k_BT/q) \ln(N_C/N_D)$): Voltage difference between conduction band minimum and Fermi level of n-Si, $V_{FB}$: Flat-band voltage, $\phi_g$($=4.50$ eV \cite{Wg}): Work-function of undoped graphene, $\Delta E_F$: Energy difference between Dirac point and Fermi level of graphene, $k_B$: Boltzmann constant, $N_C$($=2.82\times10^{19}$ cm$^{-3}$ \cite{Si}): Conduction band effective density of states, $\epsilon_0$: Vacuum permittivity, $\epsilon_s$($=11.7$ \cite{Si}): Dielectric constant of Si, $\epsilon_{ox}$($=3.9$ \cite{SiO2}): Dielectric constant of SiO$_2$, $N_D$($=3\times10^{15}$ cm$^{-3}$): Donor concentration of n-Si, $q$: Elementary charge, $\hbar$: Reduced Planck constant, $v_F$($=10^6$ m/s \cite{Wg}): Fermi velocity of graphene.

All of known constants used in the calculation are obtained from literature \cite{DFT,Wg,Al,Pt,SiO2,Si}. At the flat-band voltages, depletion or accumulation charge densities of both MOS and MGOS junctions are zero. Hence, it is reasonable to assume the charge density distributions across junctions as shown in Fig. \ref{SF1}. First, in the MOS junction, the electric field $E(x)$ and potential $U(x)$ can be calculated by using the Gauss' law with the charge distributions shown in Fig. \ref{SF1}(a) as follows.
\begin{gather}
E(x)=
\begin{cases}
\frac {1} {\epsilon_{ox} \epsilon_0} Q_m \qquad\qquad~ & (0\le x<t_{ox})
\\0 & (t_{ox}\le x)
\end{cases}
\\U(x)=
\begin{cases}
V_{FB}-\frac {1} {\epsilon_{ox} \epsilon_0} Q_m x & (0\le x< t_{ox})
\\V_{FB} - \frac {1} {\epsilon_{ox} \epsilon_0} Q_m t_{ox} & (t_{ox}\le x)
\end{cases}
\end{gather}

Also, the net charge density across the junction should be zero because the electric fields at both ends of junction are zero. That is, $Q_m+Q_{ox}=0$. By the way, the electrostatic potential change across the junction can be reflected on the potential at $x=t_{ox}$, $U(t_{ox})=\phi_m-\chi_s-V_n$. By using the potential at $x=t_{ox}$ as the boundary condition of eq. 2, we can obtain the relation between $Q_{ox}$ and $V_{FB}$.
\begin{gather}
\phi_m-\chi_s-V_n = V_{FB} - \frac {1} {\epsilon_{ox} \epsilon_0} Q_m t_{ox} \nonumber
\\Q_{ox}=\frac {\epsilon_{ox} \epsilon_0} {t_{ox}} (\phi_m-\chi_s-V_n-V_{FB})
\end{gather}

Based on the $V_{FB}$ extracted from the C-V curves of MOS junction, the $Q_{ox}$ for the substrate used in this experiment is found to be $9.02\pm1.02\times10^{-8}$ C/cm$^{2}$. Here, it is reasonable to assume that the $Q_{ox}$ obtained for MOS junction is identical to that of MGOS junction because both junctions are fabricated on the same SiO$_2$/Si substrate and differentiated only by the existence of graphene interlayer. Hence, we used $Q_{ox}=9.02\pm1.02\times10^{-8}$ C/cm$^{2}$ also to calculate the $E(x)$ and $U(x)$ for MGOS junctions with the charge distributions shown in Fig. \ref{SF1}(b).
\begin{gather}
E(x)=
\begin{cases}
\frac {1} {\epsilon_0} Q_m & (0\le x<d_1)
\\ \frac {1} {\epsilon_0} (Q_m+Q_g+Q_i) & (d_1\le x<d_1+d_2)
\\ \frac {1} {\epsilon_{ox} \epsilon_0} (Q_m+Q_g+Q_i+Q_s) \qquad\qquad~~~~~~~~~ & (d_1+d_2\le x<d_1+d_2+t_{ox})
\\ 0 & (d_1+d_2+t_{ox}\le x)
\end{cases}
\\ \nonumber
\\U(x)=
\begin{cases}
V_{FB}-\frac {1} {\epsilon_0} Q_m x & (0\le x<d_1)
\\
\\ V_{FB}-\frac {1} {\epsilon_0} Q_m d_1 -\frac {1} {\epsilon_0} (Q_m+Q_g+Q_i)(x-d_1) & (d_1\le x<d_1+d_2)
\\
\\ V_{FB}-\frac {1} {\epsilon_0} Q_m d_1 -\frac {1} {\epsilon_0} (Q_m+Q_g+Q_i) d_2
\\ -\frac {1} {\epsilon_{ox} \epsilon_0} (Q_m+Q_g+Q_i+Q_s)(x-d_1-d_2) & (d_1+d_2\le x<d_1+d_2+t_{ox})
\\
\\ V_{FB}-\frac {1} {\epsilon_0} Q_m d_1 -\frac {1} {\epsilon_0} (Q_m+Q_g+Q_i) d_2
\\ -\frac {1} {\epsilon_{ox} \epsilon_0} (Q_m+Q_g+Q_i+Q_s) t_{ox} & (d_1+d_2+t_{ox}\le x)
\end{cases}
\end{gather}
~

Similarly to MOS junction, the $Q_m$ can be calculated by using the boundary condition $U(d_1+d_2+t_{ox})=\phi_m-\chi_s-V_n$ which can be conceived in Fig. \ref{SF1}(d) and the charge neutrality condition $Q_m+Q_g+Q_i+Q_s+Q_{ox}=0$. From eq. 5,
\begin{align}
\phi_m-\chi_s-V_n&=V_{FB}-\frac {1} {\epsilon_0} Q_m d_1 -\frac {1} {\epsilon_0} (Q_m+Q_g+Q_i) d_2 -\frac {1} {\epsilon_{ox} \epsilon_0} (Q_m+Q_g+Q_i+Q_s) t_{ox} \nonumber
\\&=V_{FB}-\frac {1} {\epsilon_0} Q_m d_1 +\frac {1} {\epsilon_0} (Q_s+Q_{ox}) d_2 +\frac {1} {\epsilon_{ox} \epsilon_0} Q_{ox} t_{ox} \nonumber
\end{align}
\begin{align}
Q_m=\frac {\epsilon_0} {d_1} (V_{FB}-\phi_m+\chi_s+V_n) + \frac {d_2} {d_1} (Q_s+Q_{ox}) + \frac {t_{ox}} {d_1 \epsilon_{ox}} Q_{ox}
\end{align}

Here the trap charge density on the SiO$_2$ surface $Q_s=$ -1.12$\times$10$^{-6}$ C/cm$^2$ is obtained from the transfer curve of a graphene field effect transistor fabricated on highly p-type doped Si substrate with thermally-grown SiO$_2$ layer. $Q_s$ is considered not to change in the gate voltage range used in our experiments since the trap states are known to be $\sim$1.5 eV below from the Fermi-level of graphene \cite{Ostate}. Since the metals used in our works are known to form weak bonding with graphene enough to preserve the linear dispersion relation of graphene, the doping charge density of graphene $Q_g$ can be expressed as follows \cite{Qg}.
\begin{align}
Q_g&=\frac{q\Delta E_F |\Delta E_F|} {\pi \hbar^2 v_F^2}
\end{align}

From the flat-band diagram drawn in the Fig. \ref{SF1}(d), $\Delta E_F$ can be expressed as $V_{FB} + \phi_m - \phi_g -U(d_1)$. Hence, $Q_g$ can be obtained from $Q_m$ by using eqs. 5-7. Then, the interaction dipole charge $Q_i$ can finally be determined by using the charge neutrality condition.
\begin{align}
Q_g&=\frac{q ( V_{FB} + \phi_m - \phi_g -U(d_1) ) | V_{FB} + \phi_m - \phi_g -U(d_1) | } {\pi \hbar^2 v_F^2} \nonumber
\\&=\frac{q \Big( \phi_m - \phi_g + \frac{1} {\epsilon_0}Q_m d_1 \Big) \Big| \phi_m - \phi_g + \frac{1} {\epsilon_0}Q_m d_1 \Big| } {\pi \hbar^2 v_F^2}
\end{align}
\begin{align}
Q_i&=-Q_m-Q_s-Q_{ox}-Q_g \nonumber
\\&=-Q_s-Q_{ox}- \frac {\epsilon_0} {d_1} (V_{FB}-\phi_m+\chi_s+V_n) - \frac {d_2} {d_1} (Q_s+Q_{ox}) - \frac {t_{ox}} {d_1 \epsilon_{\rm SiO_2}} Q_{ox} - Q_g
\end{align}

In Table \ref{ST1}, the charge densities and effective work-functions of MOS and MGOS junctions obtained from the experimentally-measured $V_{FB}$, $Q_s$ and $Q_{ox}$ are listed. The wide tuning of $\phi_M^e$ of Al/SiO$_2$/Si junction with the graphene interlayer appears to originate from negative $Q_i$. The role of $Q_i$ is supported by the decrease of $\phi_M^e$ for Pt electrode where $Q_i$ turns out to be positive. These results are consistent with the previous works performed on GaAs Schottky junctions \cite{NP}.

\renewcommand{\arraystretch}{1.5}
\begin{table}[h]
\sfigone
\begin{ruledtabular}
\begin{tabular}{lcccc}
 & Al/SiO$_2$/Si & Al/G/SiO$_2$/Si & Pt/SiO$_2$/Si & Pt/G/SiO$_2$/Si \\ \hline
$V_{FB}$ (V) & -0.98 & 0.06 & 0.57 & 0.37 \\
$Q_{ox}$ (10$^{-8}$ C/cm$^{2}$) & 9.02$\pm$1.02 & 9.02$\pm$1.02 & 9.02$\pm$1.02 & 9.02$\pm$1.02 \\
$Q_m$ (10$^{-6}$ C/cm$^{2}$) & -0.09 & 1.27 & -0.09 & -2.12 \\
$Q_s$ (10$^{-6}$ C/cm$^{2}$) & - & -1.12 & - & -1.12 \\
$Q_g$ (10$^{-7}$ C/cm$^{2}$) & - & 0.37 & - & 15.44 \\
$Q_i$ (10$^{-7}$ C/cm$^{2}$) & - & -2.85 & - & 15.98 \\
$\phi_m^e$ (eV) & 3.31 & 4.35 & 4.85 & 4.66 \\ 
\end{tabular}
\caption{\label{ST1} Charge densities and effective work-functions of metal/SiO$_2$/n-Si and metal/graphene/SiO$_2$/n-Si junctions.}
\end{ruledtabular}
\end{table}
\renewcommand{\arraystretch}{1}
~

\section{The effect of Interaction dipole layer}

In order to examine the necessity of $Q_i$ for explaining the change of effective work-function, we have tried to produce the experimental outcomes just by using $d_2$ as a variable with $Q_i$ excluded. As noted previously, the other parameters can be determined from our own experiments or obtained from literature. Hence, if we substitute $Q_g=-Q_m-Q_s-Q_{ox}$ to the Eq. (7), the following quadratic equation for $Q_m$ can be obtained.
\begin{align}
-Q_m-Q_s-Q_{ox}=\frac{q \Big( \phi_m - \phi_g + \frac{1} {\epsilon_0}Q_m d_1 \Big) \Big| \phi_m - \phi_g + \frac{1} {\epsilon_0}Q_m d_1 \Big| } {\pi \hbar^2 v_F^2}
\end{align}
\begin{align}
\frac{d_1}{\epsilon_0}Q_m^2+\Big[\frac{2d_1}{\epsilon_0}(\phi_m-\phi_g)\pm\frac{\pi \hbar^2 v_F^2}{q}\Big]Q_m+(\phi_m-\phi_g)^2\pm\frac{\pi \hbar^2 v_F^2}{q}(Q_s+Q_{ox})=0
\end{align}

\renewcommand{\arraystretch}{1.5}
\begin{table}
\sfigtwo
\begin{ruledtabular}
\begin{tabular}{l|cc|cc}
& \multicolumn{2}{c|}{Al/G/SiO$_2$/Si \qquad\qquad\qquad} & \multicolumn{2}{c}{Pt/G/SiO$_2$/Si \qquad\qquad\qquad} \\ \hline
$d_2$ (\AA) & 3.9 & 5.7 & 3.9 & 5.7 \\
$V_{FB}$ calculated from eq. 11 (eV) \qquad\qquad & {\color{red}-0.15} & 0.06 & 0.37 & {\color{red}0.58}
\end{tabular}
\caption{\label{ST2} $d_2$ bearing the measured $V_{FB}$ of metal/graphene/SiO$_2$/n-Si junctions without the $Q_i$. Red-colored values of $V_{FB}$ are not consistent with the experimental values.}
\end{ruledtabular}
\end{table}
\renewcommand{\arraystretch}{1}

Here, $\pm$ sign corresponds to the case of $\Delta E_F\geq0$ and $\Delta E_F<0$, respectively. By solving second-order equation for $Q_m$, four different values of $Q_m$ can be obtained. Among them, we can choose the one that makes the sign of $Q_g$($=-Q_m-Q_s-Q_{ox}$) identical to that of $\Delta E_F$. Also, corresponding $V_{FB}$ can be calculated by substituting the $Q_m$ to eq. 6. It is reduced to
\begin{align}
V_{FB}=\frac{1}{\epsilon_o}Q_m d_1-\frac{1}{\epsilon_0}(Q_s+Q_{ox})d_2-\frac{1}{\epsilon_{ox}\epsilon_0}Q_{ox} t_{ox}+\phi_m-\chi_s-V_n
\end{align}

\begin{figure}[h]
\sfigtwo
\includegraphics{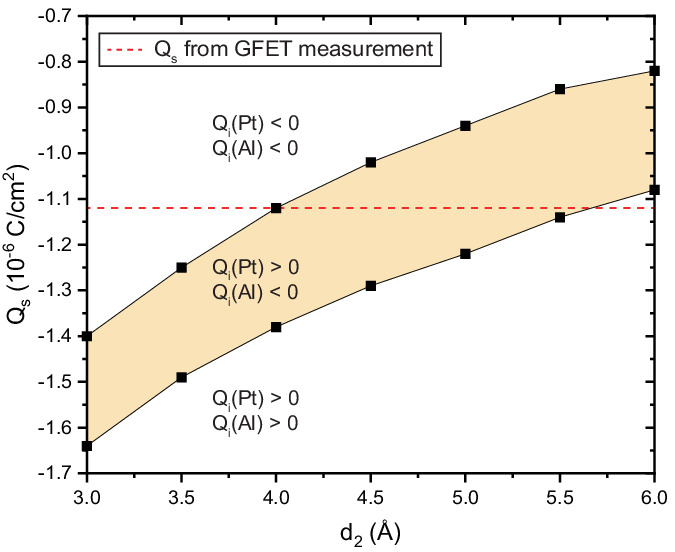}
\caption{\label{SF2} The range of $Q_s$ which makes the $Q_i<0$ for Al and $Q_i>0$ for Pt for the different $d_2$.}
\end{figure}

As mentioned previously, $V_{FB}$ for Al/graphene/SiO$_2$/Si junction is measured to be 0.06 V and that of Pt/graphene/SiO$_2$/Si junction is measured to be 0.37 V. As noticed in Table \ref{ST2}, $d_2$ needs to be assumed differently between {\it with} and {\it without} graphene interlayer to bear the experimentally-measured $V_{FB}$. However, it is not reasonable to use different values of $d_2$ because we used the same graphene transfer method for both junctions. Hence, it seems inevitable to assume the existence of $Q_i$ which is supported by the previous study \cite{NP}. In our formalism, from eq. 8, the polarity of $Q_i$ can vary depending on the value of $d_2$. In fact, our approach is NOT accompanied with any reliable experimental method to specify the value of $d_2$. This is certainly the limitation of our works. By considering this limitation, we have investigated how the polarity of $Q_i$ is related to $Q_s$ and $d_2$. In the colored area of Figure \ref{SF2}, $Q_i$ is negative for Al and positive for Pt. Here, the red dashed line indicates the $Q_s$ ($=-1.12\times10^{-6}$ C/cm$^2$) obtained from the transfer curve measured on graphene field effect transistor (GFET). As can be seen in the figure, 5 \AA ~of $d_2$ used in our electrostatic modeling seems positioned well inside the colored area where $Q_i$ is negative for Al and positive for Pt. These polarities of $Q_i$ are consistent with the previous study for the GaAs Schottky junction with the graphene interlayer \cite{NP}.

